# Improved GaInP/GaAs/GaInAs inverted metamorphic triple-junction solar cells by reduction of Zn diffusion in the top subcell

Manuel Hinojosa, Iván Lombardero, Carlos Algora and Iván García*

## Abstract

The growth of heavily doped tunnel junctions in inverted metamorphic multijunction solar cells induces a strong diffusion of Zn via a point-defects-assisted diffusion mechanism. The redistribution of Zn can compensate the n-type doping in the emitter of the GaInP top junction, degrading severely the conductivity of the whole solar cell and its conversion efficiency. This work evaluates different epitaxial growth strategies to achieve control on the [Zn] profile of an inverted metamorphic triple-junction structure, including: the reduction of the doping concentration in the tunnel junction to minimize the injection of point defects that trigger the diffusion mechanism; the use of different barrier layers to keep the injected point defects away from active layers and, finally, the minimization of Zn in the AlGaInP back-surface-field layer of the GaInP subcell. This last approach enables a high-conductivity multijunction solar cell device without redesigning the tunnel junction as well as a high electronic quality in the GaInP subcell, which shows a collection efficiency higher than 93% and an open-circuit-voltage offset of 410 mV at 1 sun irradiance. The characterization of final triple-junction devices, including quantum efficiency, electroluminescence, and light current-density-voltage curves at different irradiances, demonstrates a successful integration of all the subcell and tunnel junction components. This way, final solar cells with peak efficiencies exceeding 40 % at ~500 suns are demonstrated, despite using doping levels as low as $1·10^{17}$ cm$^{-3}$ in the AlGaInP:Zn back-surface-field of the GaInP subcell and using non-optimized antireflective coatings.

## 1. Introduction

Inverted metamorphic multijunction (IMM) solar cells offers the highest light-to-electricity conversion efficiencies among all the multijunction solar cells (MJSC) technologies by integrating high-quality lattice-mismatched bandgap engineered materials into monolithic devices [1]–[5]. In contrast with traditional upright MJSC approaches, in n-on-p IMM structures, the different junctions are deposited from higher to lower bandgaps, and, within each junction, the growth of n-type layers precede the growth of p-type layers. This way, the deposition of the lattice-mismatched materials is left to the end of the deposition process, minimizing the propagation of threading dislocations (TD) and the density of defects in the upper, high power-producing subcells [6].

However, the inverted growth direction of these structures presents specific challenges related to the diffusion dynamics of the numerous species involved in the epitaxy. For instance, the front contact layer and the highest bandgap junctions suffer the thermal load corresponding to the epitaxy of the rest of the structure and are more susceptible from suffering diffusion of elements. The exacerbation of the diffusion processes hinders the achievement of the target doping profile in these components, which compromises the solar cell performance [7]–[9]. In this regard, we recently showed that the growth of the n-side of the tunnel junction (TJ) after the GaInP subcell of an inverted multijunction solar cell was the main responsible of exacerbating the out-diffusion of Zn from the AlGaInP:Zn back surface field (BSF) layer to the GaInP top cell absorber layer via a point-defects-assisted mechanism [10], [11].

The understanding of the interaction between the growth of some layers and the Zn diffusion processes in other layers is key to gain control on the final doping profile achieved in the multijunction structure. Briefly, the point-defects-assisted mechanism works in the following way. During the growth of a heavily n-type doped layer (such as the GaAs cathode of a TJ), a high number of group-III interstitial Ga is injected at the growth surface and they rapidly diffuse into the already grown structure [12]. These injected point defects interact with acceptor Zn species (electrically active), located in the group-III sublattice of layers such as the AlGaInP:Zn BSF layer of the GaInP top junction. By means of the kick-out mechanism, Zn species are moved out of its substitutional location, generating interstitial Zn and reducing the original p-type electrical doping concentration in the layer. Once in the interstitial position, the diffusivity of Zn is orders of magnitude higher than in a substitutional positions [13]. As a result, Zn rapidly spreads out across the semiconductor structure, deviating the resulting doping profile from the intended one.

Zn is the most common acceptor dopant used in phosphide materials grown by metal-organic vapour-phase epitaxy (MOVPE). Therefore, the development of strategies to control Zn diffusion becomes interesting not only from the perspective of III-V solar cells but also for any MOVPE-grown optoelectronic device. Besides, the presence of heavily n-type layers is typically required in most optoelectronic devices, which challenges the achievement of abrupt profiles of Zn in MOVPE-grown structures due to the diffusion enhancement they induce, as explained above. Hence, although this work focuses on IMM triple-junction (IMM-3J) solar cells, the pathways proposed herein to reduce the

Zn diffusion can be extrapolated to any other optoelectronic devices requiring the growth of heavily n-type doped layers after Zn-doped layers.

In the case of MJSC, the anomalous diffusion of Zn has been demonstrated to degrade their performance in a great variety of ways; for instance, by introducing internal resistive barriers or by spoiling the passivation of the active junctions [1], [14]–[16]. For instance, in the GaInP/GaAs/GaInAs IMM-3J structure developed at our laboratory, we observed that the diffusion of Zn (p-type dopant in GaInP) was responsible of reducing severely the conductivity of the emitter of the GaInP top junction, by compensating its original n-type doping concentration. The achievement of an appropriate doping level in the emitter of the top junction is crucial to ensure a high conductivity of majority carriers to reduce the resistance to lateral current spreading towards the front grid metal fingers. Therefore, a strong diffusion of Zn in the top GaInP subcell absorber challenges the achievement of a high conductivity in the emitter, spoiling the global conductivity of the solar cell, increasing the series resistance electrical losses and, finally, reducing the conversion efficiency [17].

The strategies proposed in this work to reduce the amount of diffusing Zn focus on acting on the different stages of the point-defects-assisted diffusion mechanism. Mainly: 1) the reduction of the doping concentration in the TJ to minimize the injection of point defects; 2) the use of different barrier layers to keep the injected point defects away from GaInP top cell active layers and, finally, 3) the minimization of Zn in the back-surface-field (BSF) AlGaInP layer of the GaInP subcell. Despite most solar cell designs rely on highly doped window and BSF layers to confine efficiently the minority carriers, we found out that the most effective approach was the use of low [Zn] in the AlGaInP:Zn BSF of the GaInP junction. It is noteworthy that, in our IMM-3J structure, we use rear heterojunction (RHJ) GaInP solar cells to take advantage of the higher radiative efficiencies as well as the higher emitter conductivities in comparison with traditional front homojunction (FJ) architectures [18], [19].

The possibility of reducing the amount of Zn in the GaInP subcells has been instrumental to implement devices without having to readjust the design of the TJ to minimize the diffusion effect. By applying this approach, we demonstrate that the reduction of Zn in the structure enables 1) a successful integration of all the subcells of the IMM-3J solar cell, with collection efficiencies above 93% and open-circuit-voltage offsets ($W_{oc}$) below 0.45 V at 1-sun irradiance in all of them and 2) a high conductivity in IMM-3J solar cells, which results in conversion efficiencies exceeding 40% at ~500 suns irradiances.

## 2. Experimental

All samples were grown on GaAs substrates with a 2(111)$_B$ miscut in a horizontal low-pressure MOVPE reactor (AIX200/4). The growth conditions of the GaInP compositionally graded buffer (CGB) layer that bridges the 2% lattice-mismatch between the GaAs substrate and the 1 eV GaInAs bottom subcell are described in [20]. The precursors used were AsH$_3$, PH$_3$ for group-V, TMGa and TMIn for group-III and DTBSi, DETe and DMZn for dopant elements. The free-carrier and Zn concentration depth profiles were measured by electrochemical capacitance-voltage (ECV) and secondary ion mass spectroscopy (SIMS) scans, respectively. All solar cells were fabricated using the IMM solar cell fabrication process [21]. The active area is 0.09 cm$^2$ and the front grid pattern consists of an inverted square geometry with a nominal shadowing factor of ~4%, designed to have the maximum efficiency at 1000 suns of AM1.5d solar spectrum. The AM1.5d used in the work is normalized to 1000 W/m$^2$ at 1 sun to ease the comparison of 1 sun photocurrents between the direct and the global spectra. Front contacts are based on a Pd/Ge/Ti metal system with specific semiconductor/metal resistance of $10^{-6}$ Ω·cm$^2$ and a metal sheet resistance around 30 mΩ/sq [22]. Pd/Ge/Ti/Pd/Al is deposited using a multi-pocket electron beam evaporator. An ex-situ thermal annealing of the metallization is conducted using a rapid thermal annealing (RTA) tool, at temperatures compatible with our bonding material [23]. Specific contact resistance, contact layer sheet resistance and emitter sheet resistance ($R_{she}$) are obtained using the transmission line method (TLM). Solar cell characterization includes external quantum efficiency (EQE) and reflectance (R), carried out using a custom-made system based on a Xe lamp and grating monochromator. The internal quantum efficiency (IQE) is calculated from the EQE and the R as IQE=EQE/(1-R). Dark and light current-density-voltage (J-V) curves were taken using a Keithley 2602A instrument and the light source was a Xe-lamp based solar simulator using high intensity LED light sources of 530 nm, 740 nm and 940 nm and calibrated reference solar cells to reproduce the AM1.5g spectrum. Electroluminescence (EL) measurements were taken using a Keithley 2602A instrument for current bias and a calibrated spectroradiometer for light detection. The extraction of J-V curves from EL measurements is made following methods described elsewhere [24], [25]. The open-circuit voltage offset ($W_{oc}$) is obtained from the bandgap ($E_g$) and the open-circuit voltage ($V_{oc}$) as $W_{oc} = E_g - V_{oc}$. The solar cell devices used for EQE and R measurements have front contacts without grid to eliminate shadowing on the measurement. Concentration J-V curves were obtained using a custom made, flash-lamp based setup that uses reference (isotype) solar cells to calibrate the spectral irradiance.

# 3. Reduction of Zn diffusion in the GaInP top subcell

Fig. 1 shows the schematic of the IMM-3J solar cell structure. As described above, the analysis presented here focuses on the out-diffusion of Zn from the AlGaInP:Zn BSF towards the GaInP absorber occurring during the epitaxy growth, so it is important to identify the stages comprised in the mechanism that rules the Zn diffusion process and the layers involved. The point defects species driving the diffusion mechanism are represented in Fig.1, right. Let us focus on what happens during the deposition of the tunnel junction 1. First, during the growth of the highly doped GaAs cathode, a high quantity of group-III interstitials is injected into the already grown structure ($I_{Ga}$). It is remarkable that the quantity of $I_{Ga}$ is dependent on the doping concentration of the $n^{++}$-GaAs cathode [12], [13], [26]. Second, these $I_{Ga}$ point defects diffuse into the already grown structure, reaching the AlGaInP:Zn layer. There, $I_{Ga}$ kicks out substitutional $Zn_{Ga}$ (acceptor) species, pushing electrically active Zn species to interstitial positions $I_{Zn}$. Note that this reaction reduces the p-type doping concentration in the AlGaInP:Zn BSF layer. Third, $I_{Zn}$ propagates along the GaInP absorber where the reverse reaction takes place. By this way, the diffusing Zn gets back to the group-III lattice ($Zn_{Ga}$) and generates local holes that counterbalance the n-type doping of the GaInP absorber. This sequence of events modifies the doping profile and degrades the global device conductivity, as explained before.

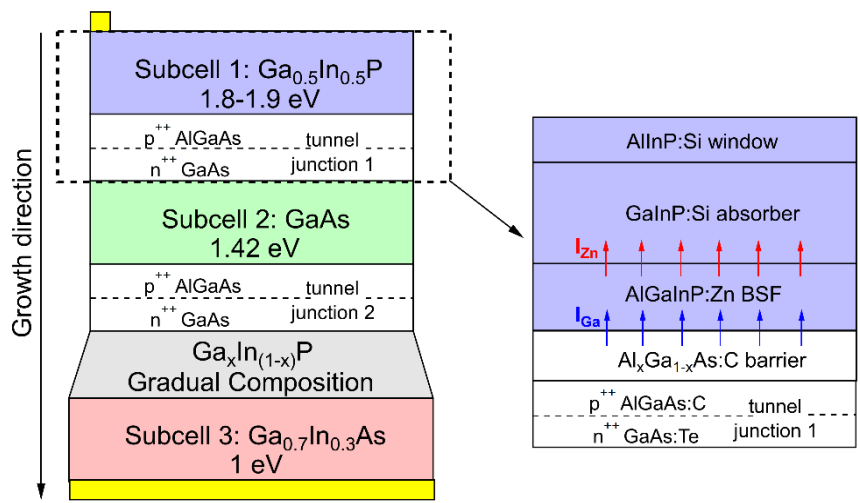

Figure 1: Schematic of the GaInP/GaAs/GaInAs IMM-3J solar cell developed in this work. In order to illustrate the Zn diffusion mechanism, we show the layers structure corresponding to the RHJ GaInP (subcell 1) plus the TJ. The propagation of interstitial Zn ($I_{Zn}$), from the AlGaInP:BSF to the GaInP absorber is represented by red arrows whereas the injection of interstitial Ga ($I_{Ga}$), from the TJ to the BSF is depicted by blue arrows. The intended doping levels in the GaInP subcell components are: [Si] = 1·10$^{18}$ cm$^{-3}$ (AlInP window), [Si] = 5·10$^{17}$ cm$^{-3}$

(GaInP absorber), [Zn] = $1\cdot10^{18}$ cm$^{-3}$ (AlGaInP BSF), [C] > $1\cdot10^{19}$ cm$^{-3}$ (AlGaAs barrier and p-side layer of the tunnel junction) and [Te]> > $1\cdot10^{19}$ cm$^{-3}$ (GaAs n-side layer of the tunnel junction). However, Zn diffusion is expected to modify the doping profile and therefore these doping levels.

The control of Zn diffusion can be addressed by acting on any stage of this sequential process. For instance, since $I_{Ga}$ is necessary to trigger the kick-out of $Zn_{Ga}$, minimizing the amount of $I_{Ga}$ would contribute to mitigate the diffusion mechanism. Thus, the reduction of the n-type doping (tellurium is used as n-type dopant) of the tunnel junction 1 can reduce the injection of $I_{Ga}$ thereby preventing the diffusion sequence. To evaluate the effect of the tunnel junction doping on the Zn diffusion in this multilayer structure, we carried out experiments on test samples that mimic the structure and growth arrangement of the GaInP junction plus the tunnel junction layers (a schematic of the structures is presented in the corresponding graphs). Note that all samples evaluated in Fig. 2 are identical, except for the [Te] in the n$^{++}$-GaAs cathode layer which ranges from $8\cdot10^{18}$ to $3\cdot10^{19}$ cm$^{-3}$. A subsequent bake in the MOVPE reactor is performed in all the test structures to reproduce the thermal load of a full MJSC structure (60 minutes at 675 °C in AsH$_3$/H$_2$ ambient).

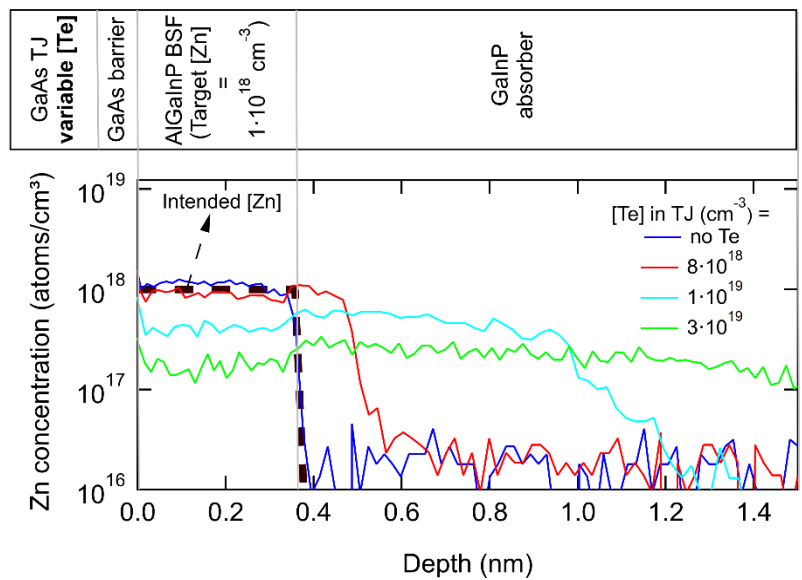

Figure 2: [Zn] depth profile measured by SIMS along the phosphide layers of test structures that mimic the GaInP top subcell of an IMM-3J structure with variable [Te] in the GaAs cathode: no doping, $8\cdot10^{18}$, $1\cdot10^{19}$ and $3\cdot10^{19}$ cm$^{-3}$. An ex-situ MOVPE bake at 675 °C for 60 minutes is performed in all samples to reproduce the growth of a multijunction structure.

Fig. 2 shows the SIMS [Zn] profile along the AlGaInP/GaInP layers of the structures with variable doping in the tunnel junction. It can be clearly perceived that the diffusion intensity is closely related to the GaAs cathode doping, becoming almost negligible in the absence of [Te]. Unfortunately, it is well-known that the performance of the TJ depends strongly on the doping level achieved in both the cathode and the anode layers. Doping levels below $1\cdot10^{19}$ cm$^{-3}$ in similar multijunction devices fabricated in our laboratory have shown series resistance problems at low irradiances. Thus, although the reduction of the GaAs doping level enables a minimized diffusion during the epitaxy, it would also spoil the tunnel junction performance and compromise the electrical interconnection of the GaInP and the GaAs subcells. Therefore, it can be concluded that, although the reduction of [Te] represents an effective way to get effective control upon the Zn profile, it becomes inadequate for the subcell interconnection.

An alternative pathway to reduce diffusion whilst keeping a high n-type doping in the TJ cathode ([Te] = $3\cdot10^{19}$ cm$^{-3}$) consists of including barriers to keep injected point defects away from the AlGaInP:Zn layer. Hence, this strategy does not pursue to reduce the injection of point defects but trapping them into inactive layers before reaching the Zn-doped layers. The tunnel junction barrier layers, adjacent to the anode and the cathode in typical III-V MJSC structures, can be contemplated for this purpose [27], [28]. If the $I_{Ga}$ species injected from the tunnel junction are kept away from Zn-doped layer, the kick out reaction can be prevented, enabling a successful confinement of electrically active Zn in the BSF layer. To accomplish this, we have investigated different barrier layers in a set of test structures (Fig. 1, right + bake), nominally identical except for the materials and thicknesses used in the barriers. This way, the differences detected in the Zn profiles (and, in the same way, in the hole profiles) can be directly attributed to the efficiency of barriers stopping point defects.

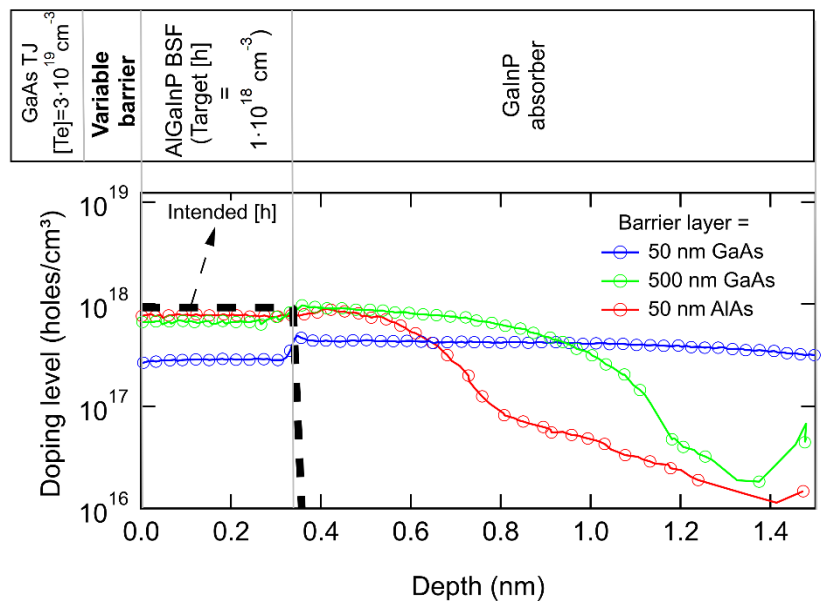

Figure 3: Hole concentration [h] depth profile along the phosphide layers of test structures that mimic the GaInP top subcell of an IMM-3J with a [Te]= 3·10$^{19}$ cm$^{-3}$ in the GaAs cathode using several barriers: 50 nm GaAs, 500 nm GaAs and 50 nm AlAs. An ex-situ MOVPE bake at 675 °C for 60 minutes is performed in all samples to emulate the growth of a multijunction structure.

The ECV profiles along the phosphide layers of samples employing a 50 nm GaAs, a 500 nm GaAs and a 50 nm AlAs barrier layers are shown in Fig. 3. We use GaAs and AlAs barrier to cover the extreme cases of AlGaAs alloys with variable aluminum content. It is noteworthy that this characterization technique does not provide the Zn concentration profile but the free-carrier concentration (holes in this case), which can be correlated to the concentration of Zn. A high doping level in the TJ cathode ([Te] = 3·10$^{19}$ cm$^{-3}$) is used in these samples. As expected, the structure with a 50 nm GaAs barrier suffers a strong Zn diffusion. However, the efficiency of the barrier to stop the point defects clearly improves by increasing its thickness (500 nm GaAs) or, better, by using an AlAs layer (50 nm AlAs). These results suggest that the Zn diffusion could be effectively addressed by optimizing the Al content and the thickness of an AlGaAs barrier between the tunnel junction 1 and the GaInP junction.

The incorporation of alternative materials in already optimized tunnel junctions can be troublesome because the use of different bandgaps, electronic affinities, and doping levels can easily introduce resistive barriers. This way, it is worth implementing structures that do not require readjusting the TJ design (neither the doping level of its components nor the barrier layers) to ensure a high solar cell performance [27]. Besides, paradoxically, the action of some point defects injected by the TJ has been demonstrated to be very beneficial for the electronic quality of the GaInP junction by boosting its internal radiative efficiency, voltage and quantum efficiency [29]. This leaves us with the third approach proposed to explore: not to limit the injection and diffusion of point defects but reducing the Zn concentration to minimize the impact of its diffusion. For this, we investigate the reduction of Zn in the AlGaInP BSF, as main source of Zn diffusion, to mitigate the n-type doping compensation in the GaInP absorber.

In this case, we use full IMM-3J structures for the study (so we can fabricate devices from these structures). In particular, we use IMM-3J structures employing a GaInP RHJ top subcell with an n-type doping level in the absorber [Si]~4·10$^{17}$ cm$^{-3}$ (IMM-3J A and IMM-3J B). Both structures are identical except for targeting different [Zn] in the AlGaInP:Zn BSF (i.e. the growth conditions are kept constant except for the DMZn injected during the growth of the AlGaInP BSF). In IMM-3J A, the nominal [Zn] in the AlGaInP:Zn BSF is 7·10$^{17}$ cm$^{-3}$ whereas in the IMM-3J B, the target [Zn] is 2·10$^{17}$ cm$^{-3}$, as measured in calibration samples. Fig. 4 shows the ECV profile along the phosphide layers of the top junction of these IMM-3J structures. The nominal sheet emitter resistance $R_{she}$ with the target doping profile, represented with red and blue dashed lines in Fig 4, is 150 Ω/sq, as measured in GaInP single junction solar cell structures with no Zn diffusion.

In IMM-3J A (Fig. 4, top), the acceptor [Zn] in the BSF drops a factor of 7 with respect to the value obtained in single junction structures, from $7·10^{17}$ (nominal value and doping achieved in single junction GaInP solar cells) to $1·10^{17}$ cm$^{-3}$. In this case, the diffusing Zn spreads along the GaInP absorber, leading to a fully compensated p-type GaInP region in the middle of the absorber layer. The free-electron concentration [n] in the rest of the GaInP layer gets partially compensated, reducing the effective doping level down to $7·10^{16}$-$2·10^{17}$ cm$^{-3}$, which significantly degrades the lateral conductivity in the emitter ($R_{she}$ = 1230 Ω/sq). Note that the resulting doping profile breaks the RHJ electronic structure, because the fully compensated region displaces the *pn* junction position away from the AlGaInP/GaInP interface. On the other hand, the ECV profile of IMM-3J B (Figure 4, bottom) shows that the diffusion is effectively stopped in the middle of the absorber layer, so both the fully compensated and the partially compensated regions are clearly reduced in thickness. The resulting doping level yields a drastically lower $R_{she}$ about 300 Ω/sq, which is appropriate for the operation at high concentrations, since the nominal value used for the design of the front metal grid for these concentrator cells is 450 Ω/sq.

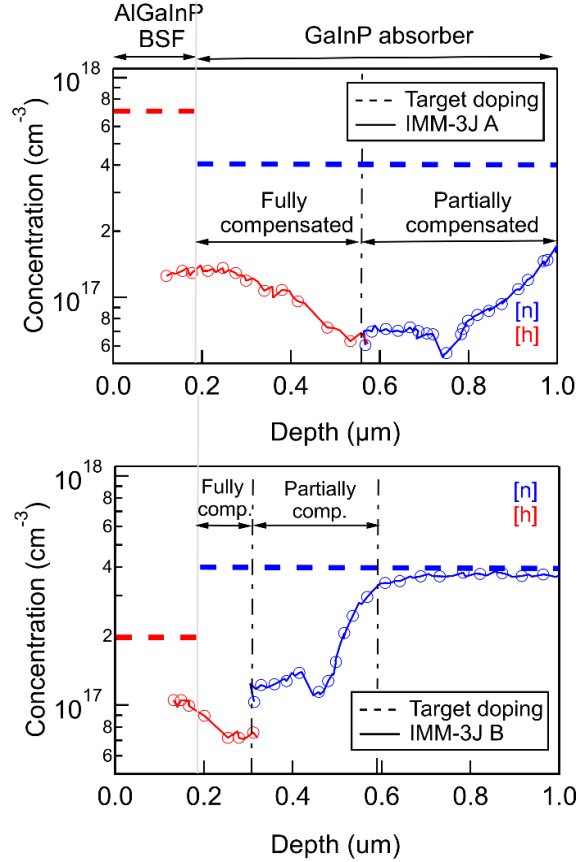

Figure 4: Doping profiles along the AlGaInP/GaInP layers of the GaInP RHJ subcell of two IMM-3J designs using different target [Zn] in the BSF: $7 \cdot 10^{17}$ cm$^{-3}$ in IMM-3J A (top) and $2 \cdot 10^{17}$ cm$^{-3}$ in IMM-3J B (bottom).

Regardless the difference in the doping profile, devices fabricated with IMM-3J A and IMM-3J B show almost identical IQE and I-V curves, which rules out a significant degradation on the recombination parameters in the GaInP subcell. The solar cell performance of IMM-3J B is characterized in next section, being the IQE and I-V curves of IMM-3J A very similar. However, the contrast in the $R_{she}$ induced by differences in the Zn profile, has a strong impact on the global conductivity of the fabricated devices. Fig. 5, top, shows the light J-V curves at different concentrations of both devices. In Fig. 5, bottom, it is shown the fill factor (FF) plotted against the short-circuit-current ($J_{sc}$). It can be clearly seen how IMM-3J B presents a significantly improved concentration response thanks to its better conductivity. By this way, FF over 86% can be obtained at irradiances exceeding 400 suns by minimizing the diffusion of Zn in the absorber of the top junction. The modeled global resistance in the IMM-3J B device is ~10 mΩ·cm$^2$ (prior to the tunnel junction failure, which occurs at more than 500 suns irradiances, because the peak current $J_{peak}$ is exceeded). This global resistance results a very suitable value for concentrator solar cells.

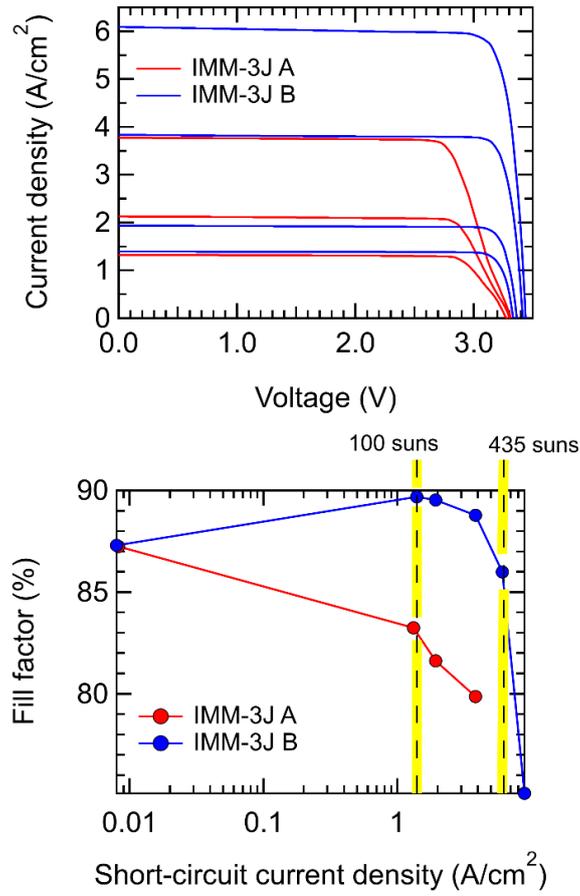

Figure 5: Comparison of the concentration response of IMM-3J A and IMM-3J B devices with notable differences in the topcell $R_{she}$ (1230 Ω/sq and 300 Ω/sq, respectively) because of Zn diffusion. Top: Light J-V at different irradiances. Bottom: Evolution of the fill factor at different short circuit current levels. The short-circuit current densities corresponding to 100 and 435 suns are indicated (considering $J_{sc}(1x)$= 14 mA/cm$^2$).

# 4. Characterization of IMM-3J solar cells with reduced Zn diffusion

Traditional III-V solar cell structures usually target high doping levels (>1·10$^{18}$ cm$^{-3}$) as well as higher bandgap materials in the confinement layers (window and BSFs) to guarantee an efficient passivation of the front and rear interfaces of the active subcells. Thus, the low [h] attained in the AlGaInP BSF of the GaInP top junction of the IMM-3J B (~1·10$^{17}$ cm$^{-3}$, as seen in the ECV profile shown in Fig.4, bottom), is significantly lower than the pursued in standard designs. Therefore, despite we have demonstrated a drastically improved FF in the design using a low doping in the BSF, concerns about the impact of this approach on other performance parameters of the IMM-3J are raised.

In this section we present the characterization of devices fabricated from IMM-3J B, i.e., with reduced Zn in the GaInP absorber (from now on, IMM-3J). First, we use the EQE and the EL techniques to estimate the $J_{sc}$ and dark J(V) characteristics of the individual GaInP, GaAs and GaInAs subcells [21], [22], [29]. These measurements are useful to evaluate the individual performance of the GaInP junction once integrated in the IMM device. Then, we present 1-sun and concentration light J-V curves of the IMM-3J device.

Model IMM-3J structures are used, which do not pursue achieving optimum efficiencies (several parameters, such as the front mask, the top cell and bottom cell bandgaps and the anti-reflection coating ARC are not still optimized at present time). The bandgaps at 25 °C extracted by EL measurements are 1.83, 1.42 and 1.00 eV. The thicknesses of the GaInP, GaAs and GaInAs absorbers (825 nm, 3000 nm and 2000 nm, respectively) have been designed aiming to achieve current matching under the AM1.5d spectrum, for the material bandgaps obtained. The GaInP and the GaInAs subcells are optically thin, to attain subcell current matching. Remarkably, due to its excessively low bandgap, a thickness of 2000 nm in the GaInAs subcell is sufficient to generate more photocurrent than in the GaAs subcell, which is the limiting subcell (especially considering the mirror effect of the back gold metallization).

Fig. 6 shows the IQE of the GaInP, GaAs and GaInAs subcells. The luminescence coupling between the junctions has been corrected as described elsewhere [30]. The modelled collection efficiencies are 93%, 97% and 94%, which are very similar to those achieved in GaInP, GaAs and GaInAs single junction solar cells fabricated at our laboratory, despite the doping in the AlGaInP:Zn BSF of the single junction GaInP solar cell is one order-of-magnitude higher (around $1 \cdot 10^{18}$ cm$^{-3}$). The collection efficiency of a solar cell indicates the number of collected minority carriers (and therefore contributing to generate photocurrent) per each photon absorbed in the absorber layer (collection efficiency = collected carriers/absorbed photons). Hence, it gives insight on how good the solar cell is generating photocurrent regardless the absorber thickness or the internal optics of the solar cell multilayer stack. The absorption at each layer of the multijunction stack is calculated by using the transfer matrix method [31]. Therefore, the high collection efficiency achieved in the integrated top subcell demonstrates that the low doping level attained in the AlGaInP layer of the IMM-3J does not significantly affect the collection efficiency of the GaInP junction.

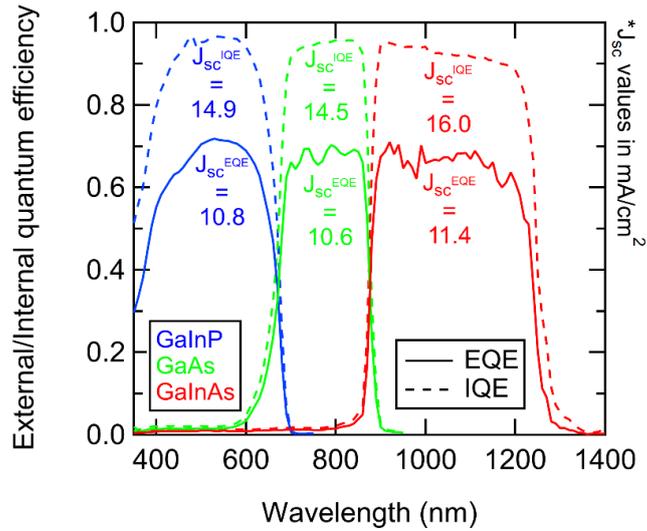

Figure 6: External/Internal quantum efficiency of the individual subcells of the IMM-3J solar cell. The modelled collection efficiencies are 93%, 97% and 94% for the GaInP, GaAs and GaInAs subcells, respectively. The measured device does not have ARC. The values of the photocurrent densities (mA/cm$^2$) calculated from integrating the EQE and the IQE of the different subcells with respect to the AM1.5d normalized to 1000 W/m$^2$ are included.

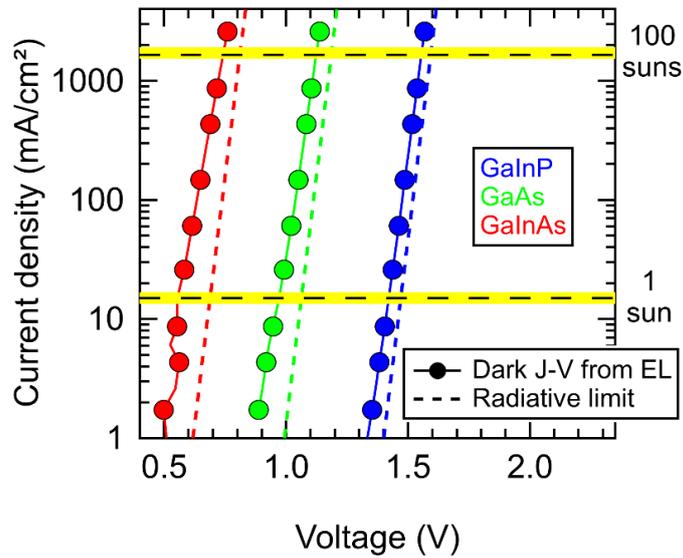

Figure 7: Dark J-V curve of the internal GaInP, GaAs and GaInAs subcells in the IMM-3J obtained by using the EL technique. The modelled curves in the radiative limit are included (note that this modelling consider the effects of the internal optics in the management of internally emitted photons). The $\eta_{ext}/\eta_{int}$ values obtained at 1 sun current densities are: 1.2/51.9 %, 0.2/25.7% and 0.1/5.5% for the GaInP, GaAs and GaInAs subcells. The bandgap-voltage offset $W_{oc}$ values of the internal subcells are 410 mV (GaInP), 455 mV (GaAs) and 440 mV (GaInAs), respectively. In the world-record efficiency IMM-3J, the $W_{oc}$ values are 410 mV (GaInP), 350 mV (GaAs) and 350 mV (GaInAs) [5].

The individual subcell dark J-V curves (GaInP, GaAs and GaInAs) extracted from the EL measurements are shown in Fig. 7. The modelled J-V curves of these junctions in the radiative limit are included, as calculated by means of the reciprocal relationship established between the EL, the EQE, the voltage and the internal optics of the multijunction stack [32]–[34]. The parameters that we use to evaluate the effects of the internal optics on the solar cell luminescence are $P_{abs}$ and $P_{esc}$, which determine the probability of an internally emitted photon to be either absorbed in the active layers of the solar cell where it was emitted (promoting photon recycling) or to escape outside the semiconductor and contribute to the external luminescence. These parameters are averaged over the spontaneous emission energy distribution and the uniform solid angle of internal emission [34]. The optical parameters calculated for the multijunction stack (average $P_{abs}$ and $P_{esc}$ of internally emitted photons) are: 1.5 % and 62.1 % for the GaInP subcell; 0.7 % and 86.1 % for the GaAs subcell and, finally, 1.2 % and 87.7 % for the GaInAs subcell. Note that these optical values determine the maximum external radiative efficiency ($\eta_{ext}$) for an ideal material quality (internal radiative efficiency $\eta_{int}$ equal to 100%), because the internal optics affects the management of internally emitted photons, which has a direct impact on the J(V), especially in the radiative limit [35]. The current density levels corresponding to 1 sun and 100 suns irradiances are highlighted in the graph as benchmarks (assuming $J_{sc}$(1X) =14 mA/cm$^2$).

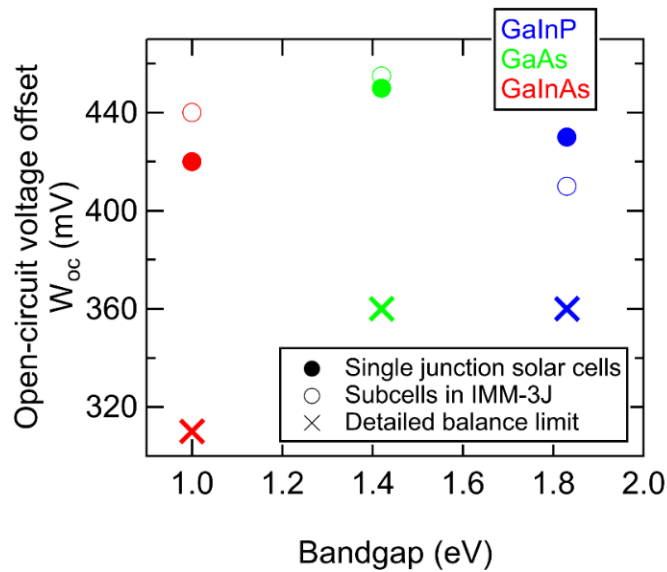

Figure 8: $W_{oc}$ of the GaInP (blue), GaAs (green) and GaInAs (red) subcells in the IMM-3J obtained by using the EL technique (empty markers) at 1 sun AM1.5d. The $W_{oc}$ values of the single junction solar cells fabricated prior to the implementation of the IMM-3J are also included in the graph (filled markers). The $W_{oc}$ values in the radiative limit of the different subcells are included, considering the modelled internal optics of the multijunction stack.

Figure 8 shows the $W_{oc}$ values of the GaInP, GaAs and GaInAs subcells integrated into the IMM-3J device at 1 sun current levels: 410 mV, 455 mV and 440 mV. The $W_{oc}$ values of the single junction solar cells fabricated prior to the implementation of the IMM-3J are also included in the graph, as well as the $W_{oc}$ of the subcells in the radiative limit, (calculated accounting for the actual internal optics of the multijunction stack). It seems that the voltage of the GaInP junction raises about 20 mV when integrated in the multijunction stack, fact that might be caused by a passivation of defects caused by the growth of the tunnel junction, as observed in similar structures [29]. In fact, the GaInP subcell shows higher material quality than the GaAs subcell in all cases. Our modelling indicates $\eta_{int}$ values at 1-sun current density of 52%, 26% and 6% for the GaInP, GaAs and GaInAs junctions. On the other hand, the metamorphic GaInAs subcell appears to suffer a subtle voltage degradation when integrated. We consider that this degradation might be induced by the growth of the 2$^{nd}$ tunnel junction prior to the deposition of the metamorphic buffer layer by means of a residual incorporation of Te or point defects, which might reduce the ordering degree of the graded GaInP layers, thereby degrading the quality of the metamorphic material [20]. In any case, it can be noticed that the $W_{oc}$ values of the subcells are found within a range of 20 mV after being incorporated into the multijunction structures, which demonstrates a successful integration of components into the IMM-3J device.

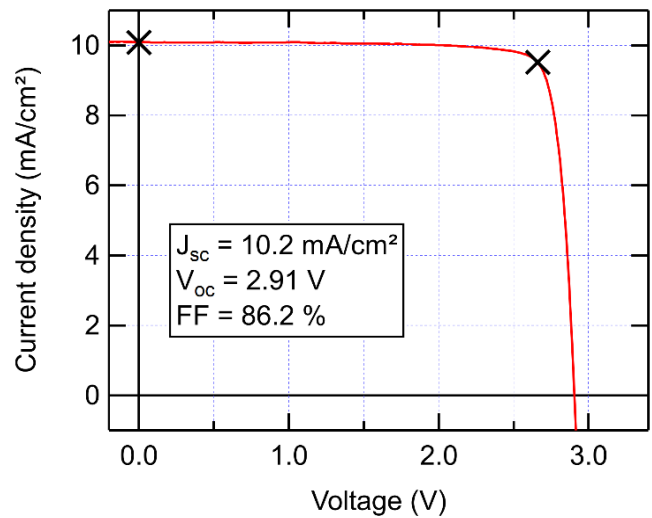

Figure 9: Light J-V curve of the IMM-3J device under 1-sun irradiance calibrated AM1.5 global spectrum (1000 W/m$^2$). The device does not have anti-reflective coating.

Fig. 9 shows the light J-V curve at 1-sun irradiance of the IMM-3J. The efficiency is low (25.59%) because no ARC is deposited. However, from the IQE and the reflectance measurements, we model potential (realistic) efficiencies exceeding 35% by depositing an adequate ARC in this device (assuming the same front-grid design, which has an excessive shadowing because is optimized for concentration applications). This modelling is made by using a distributed circuit model for the multijunction solar cell which accounts for the effects of the photocurrent mismatch between the different subcells on the global efficiency [36]. Since the total thickness of the semiconductor film is 11.5 µm, the potential power density enabled by this IMM-3J structure transferred to a flexible lightweight carrier would be as high as 5.7 kW/kg under AM1.5g 1-sun (considering a density of the semiconductor film of 5.32 g/cm$^3$ [37]). By assuming a flexible carrier such a PET foil of 25 µm, with a density of 1.38 g/cm$^3$, the potential power density would be 3.5 kW/kg. Nonetheless, at present time we are still optimizing the fabrication of 1-sun devices with ARC. We are facing some difficulties related to the isolation of devices during the fabrication of devices with ARC. However, these results aid to demonstrate a good 1 sun performance of a 3J IMM even though having reduced drastically the quantity of Zn used. Fig. 10, top shows the EQE and R of IMM-3J devices with a preliminary ARC. The one sun measurement is not presented because these preliminary devices (with ARC and front fingers) were partially shunted due to under etching issues during the device isolation stage, which spoils the FF when the $J_{sc}$ is about 14 mA/cm$^2$ (~75%). However, this current leakage is not dramatic at concentration irradiances where $J_{sc} \gg J_{shunt}$. The $J_{sc}^{EQE}$ values of the subcells calculated from the EQE of devices without front fingers under the AM1.5d normalized to 1000 W/m$^2$ are included in the figure. They show that the middle cell is limiting the current, with a bottom cell exhibiting an excessive photocurrent. Tuning the thickness of the GaInP top cell and raising the bandgap of the GaInAs metamorphic bottom cell is required to improve the current matching and performance of the IMM-3J. Fig. 10, bottom, shows the light J-V measurement of the IMM-3J device with ARC under 513 suns. Despite the limitations introduced by the imperfect current matching, the resulting efficiency is 40.4%.

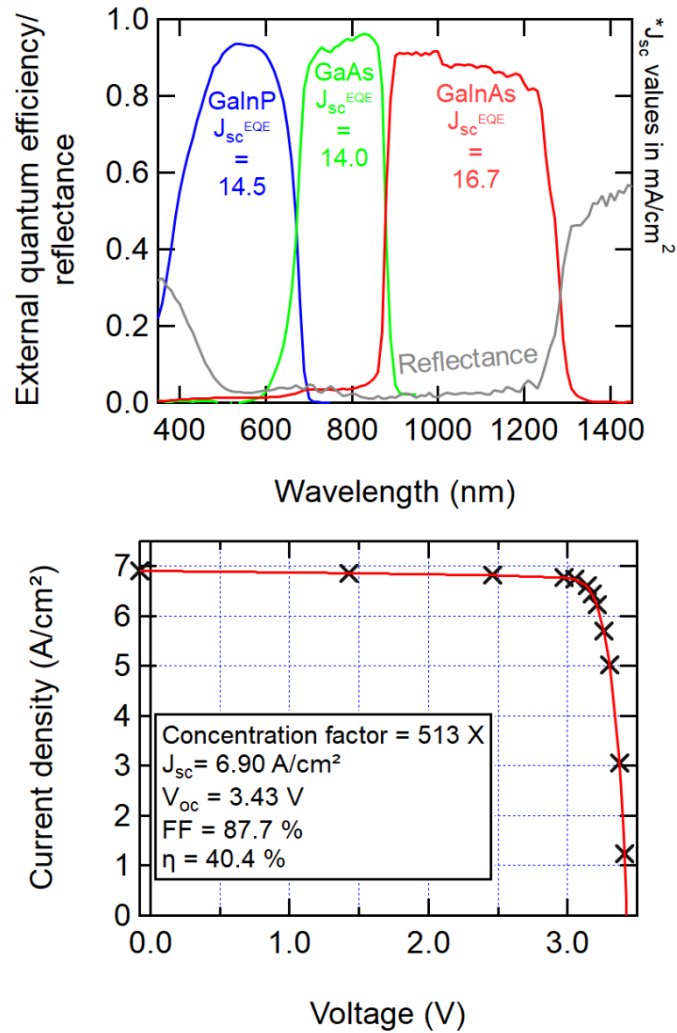

Figure 10: Top: EQE and R of IMM-3J with ARC. Bottom: Light J-V curve of the IMM-3J with ARC under 513-sun irradiance.

These results demonstrate the effectivity of the solution consisting of reducing the Zn concentration in the GaInP top cell BSF layer to minimize the impact of diffusion of this dopant during the growth of the rest of the inverted structure. Further refinements to the three-junction stack to improve current matching, solving mesa etching issues and optimizing the front grid mask and ARC are expected to enable efficiencies exceeding 36% at 1 sun and 43% under concentration.

## Conclusions

In this paper we discuss different pathways to reduce Zn diffusion in a IMM-3J, which becomes problematic because the diffused Zn compensates the doping level on the top subcell emitter and degrades the global conductivity of the device. The proposed strategies come up from the perspective of the point-defects-assisted mechanism that drives the diffusion process. The reduction of the n-type doping of the tunnel junction proves to be an effective solution to mitigate the Zn diffusion, but it might degrade the performance of the tunnel junction far beyond what could be assumed. Similarly, the use of barriers seems to be effective, but the added complexity to the structure might introduce potential energy band misalignments that can hinder the solar cell performance. Therefore, we focus on implementing a solution that does not involve substantial changes in the structure: reducing about an order-of-magnitude the Zn concentration in the GaInP top subcell BSF layer down to ~$1·10^{17}$ cm$^{-3}$. This approach succeeds in reducing the Zn diffusion and allows appropriate $R_{she}$ values of 300 Ω/sq, suitable for 1-sun and concentration applications, without compromising the performance of the integrated subcells. Concerns about possible detrimental effects of this low BSF doping on the performance of the IMM-3J are ruled out by demonstrating that the reduction of Zn in the structure enables: 1) a successful integration of all the components of the IMM-3J solar cell, with collection efficiencies above 93% and open-circuit-voltage offsets ($W_{oc}$) below 0.410 V at 1-sun current-densities in the GaInP junction and 2) a high conductivity in the resulting IMM-3J solar cells, which enables peak conversion efficiencies exceeding 40% at ~500 suns irradiances. Further refinements to the three-junction stack to improve current matching, solving mesa etching issues and optimizing the ARC are expected to enable efficiencies exceeding 36% at 1 sun and 43% under concentration.

## ACKOWLEDGMENTS

This work has been supported by the Grant PID2020-112763RB-I00 funded by the Ministerio de Ciencia e Innovación (MCIN/AEI/10.13039/501100011033), by the Project EQC2019-005701-P, funded by Agencia Estatal de Investigación (AEI/10.13039/501100011033), MCIN and European Regional Development Fund (ERDF) "A way to make Europe" and by the Universidad Politécnica de Madrid through "Ayudas para la cofinanciación de infraestructuras de I + D + i (Programa Propio)".